\documentclass[11pt,]{article}
\usepackage{authblk}
\usepackage{fullpage}
\usepackage{amssymb,amsmath}
\usepackage{titlesec}
\usepackage[utf8x]{inputenc}
\usepackage[T1]{fontenc}
\usepackage{siunitx}
\usepackage[version=3]{mhchem}
\usepackage{xr}
\usepackage{bm}
\usepackage{pdfpages}
\usepackage{graphicx}
\usepackage{dcolumn}
\usepackage{bm}
\usepackage{color}
\usepackage{tikz}
\usepackage{setspace}
\usepackage{tabularx}
\usepackage{doi}
\usepackage{url}
\usepackage[title]{appendix}

\usepackage{float}
\usepackage{mathtools}
\usepackage{braket}
\usepackage[center,font=small,labelfont=bf]{caption}

\usepackage{makecell}

\usepackage[sort&compress]{natbib}
\usepackage{setspace}
\usepackage{hyperref}
\hypersetup{colorlinks=true,
            citecolor=blue,
            linkcolor=blue}
\renewcommand\appendixname{Appendix}

\numberwithin{equation}{section}

\usepackage{graphicx}
\graphicspath{{figures/}}
\makeatletter
\def\ScaleWidthIfNeeded{%
 \ifdim\Gin@nat@width>\linewidth
    \linewidth
  \else
    \Gin@nat@width
  \fi
}
\def\ScaleHeightIfNeeded{%
  \ifdim\Gin@nat@height>0.9\textheight
    0.9\textheight
  \else
    \Gin@nat@width
  \fi
}
\makeatother
\setkeys{Gin}{width=\ScaleWidthIfNeeded,height=\ScaleHeightIfNeeded,keepaspectratio}%

\begin{document}

\title{Range residency determines how movement persistence shapes encounter rates}

\author[1]{Anudeep Surendran}
\author[2]{Luisa Ramirez}
\author[1,3,4,5]{Justin M. Calabrese}
\author[5]{William F. Fagan}
\author[1,6,7,*]{Ricardo Martinez-Garcia}

\affil[1]{\small Center for Advanced Systems Understanding (CASUS) -- Helmholtz-Zentrum Dresden-Rossendorf (HZDR), Untermarkt 20, Görlitz 02826, Germany}
\affil[2]{Institute of Developmental Biology and Neurobiology, Johannes-Gutenberg University Mainz, Mainz, Germany}
\affil[3]{Department of Geosciences, TUD Dresden University of Technology, Dresden, Germany}
\affil[4]{Department of Ecological Modelling, Helmholtz Centre for Environmental Research -- UFZ, Germany}
\affil[5]{Department of Biology, University of Maryland, USA}
\affil[6]{ICTP South American Institute for Fundamental Research \& Instituto de F\'isica Te\'orica, Universidade Estadual Paulista - UNESP, Brazil.}
\affil[7]{Department of Ecology, Institute of Biosciences, University of São Paulo, São Paulo, Brazil }
\affil[*]{Corresponding author: r.martinez-garcia@hzdr.de}

\date{\vspace{-3em}}
 
\maketitle


\begin{abstract}
Encounters between individuals link movement behavior to population-level processes such as predation and disease transmission. For many animal species, movement can be modeled as a multiscale stochastic process, dominated by directional persistence at short time scales and range residency at long time scales. Their separate effects on encounters are well understood: range residency can raise or lower encounter rates depending on home-range overlap, whereas moving with higher directional persistence systematically increases them. However, how directional persistence and range residency jointly determine encounters remains unknown. We present an analytical encounter theory for movement models that combine both features. In this framework, we derive a threshold in home-range overlap above which directional persistence diminishes, rather than enhances, encounters. At shorter time scales, attraction toward the home-range center displaces individual locations, altering encounter rates even when range residency is not measurable in movement tracks. Movement models fitted to short tracks can therefore describe trajectories accurately yet under- or overestimate the encounters derived from those trajectories, depending on home range spatial configuration. Encounter rates are a more demanding target for inference than movement parameters themselves.
\end{abstract}

\section{Introduction}\label{Sec:Introduction}

Encounters between individuals are fundamental to various ecological interactions, such as mating, predation, and disease transmission, among others \citep{Turesson_and_Brönmark_2007, Deuel_et_al_2017, Wilber_et_al_2022,talmon2025}. These individual-level encounters collectively shape broader population-level dynamics, influence population abundances and spatial distributions, community structure and stability, and ultimately play a key role in overall ecosystem functioning \citep{Mueller_and_Fagan_2008, Gil_et_al_2018, A_Surendran_2020}. One of the main factors influencing encounters is individual movement behavior \citep{Visser_and_Kiorboe_2006, Merrill_et_al_2010, Surendran2019, Martinez-Garcia_et_al_2020}. Consequently, a major challenge in ecology is to understand how empirically observed movement patterns shape encounters and, in turn, broader ecological processes \citep{Shaw_2020,hein_information_2020, Costa-Pereira_et_al_2022}.

At shorter time scales, encounters also shape how animals use space.  Mechanistic models of territory formation show that individuals' long-term space-use patterns emerge from the cumulative effects of repeated encounters with conspecific neighbors \citep{Moorcroft2013,Potts2014}. Empirically, encounters between conspecifics can drive shifts in home-range overlap \citep{Moorcroft_et_al_2006, Giuggioli2014, Fagan2024}, the defense of territories \citep{Christensen2018, Hinsch2017}, the avoidance of areas where neighbors are likely \citep{Mech2002}, and more cautious movement through encounter areas \citep{Torrez2020, Noonan2021}. Likewise, encounters with heterospecifics can drive resource partitioning \citep{Ferry2016}, shifts in activity timing \citep{moretto_free-ranging_2026}, and determine the intensity of predator-prey interactions \citep{Noonan_et_al_2023}.

Despite the ecological importance of movement-mediated encounters, existing models of encounter rates largely rely on mass-action frameworks, which assume that animals can move throughout the entire population range and that encounters can be treated as statistically independent events \citep{Hutchinson2007, Figueiredo2025}. This assumption, however, overlooks two features of real trajectories. Many organisms move within home ranges smaller than the population range \citep{Burt1943, Powell2012}, and movement at a finite velocity correlates an individual's location over time, so that the interval between two consecutive encounters depends on when the previous one occurred. 

Directional persistence is expected to always favor encounters, as less tortuous trajectories increase the effective area an individual sweeps \citep{Bartumeus_et_al_2008, Visser_and_Kiorboe_2006, Kiorboe2005}. This benefit is bounded to short time scales, before velocity correlations decay and correlated random walks converge to Brownian motion \citep{Bartumeus_et_al_2008}. At long time scales, range residency confines encounters to regions of home-range overlap and causes the encounter rate to increase or decrease relative to mass-action expectations, depending on the home-range separation \citep{Martinez-Garcia_et_al_2020, Noonan2021}. These two results describe encounters at different time scales, and whether the short-time benefit of directional persistence survives into the long-time regime in which range residency operates remains unknown.

Animals display both behaviors in the same trajectory. Analyses of long, high-resolution tracking data \citep{Hebblewhite2010,Foley2020} show that movement can be often understood as a multiscale stochastic process \citep{Fleming_et_al_2014}.
At short time scales, animal movement is frequently dominated by persistent, highly autocorrelated segments, but the same movement processes converge to a saturating mean squared displacement at long time scales. This phenomenological richness reflects the complexity of animal behaviors, with more persistent movements corresponding to directional travel and foraging excursions, and the saturation in the mean-squared displacement indicating the existence of home ranges \citep{Fleming_et_al_2014}. Recent analyses of large multispecies animal-movement datasets further indicate that these two key movement features are prevalent across a broad range of animal species \citep{Noonan2019, Fagan2025}.

These developments originated in the statistical analysis of animal movement tracks \citep{Fleming_et_al_2014, Fleming_etal_2014b, Fleming_et_al_2015} and have outpaced theoretical progress on encounter rates, with researchers only recently beginning to explore how features like range residency and correlated movements contribute to encounter rates \citep{das_misconceptions_2023,Martinez-Garcia_et_al_2020,GarciaFigueiredoRoads2025,hein_information_2020}. This recent progress toward a more empirically grounded view of ecological encounter rates has been fostered by the application of well-established condensed-matter-physics techniques \citep{Kenkre2021,Montroll,Sarvaharman2023particle}, which have enabled the analysis of encounters as reaction-diffusion stochastic processes \citep{Gurarie2012, Figueiredo2025}. Despite a few promising recent developments \citep{Martinez-Garcia_et_al_2020,hein_information_2020,das_misconceptions_2023,GarciaFigueiredoRoads2025}, a more general view of ecological encounter rates that incorporates movement as a multiscale stochastic process has remained elusive.

To tackle this issue, we introduce a theoretical framework for animal encounters that captures the complexities of individual movement across multiple time scales, combining short-term autocorrelated velocity with long-term range residency \citep{Fleming_et_al_2014}. We then demonstrate the rich behaviors in encounter dynamics that arise from the interplay among these movement features. This framework makes the consequences of multiscale movement for encounters analytically tractable, providing the theoretical tools to connect patterns of individual movement to the population-level processes they govern.

\section{Theoretical framework}

\subsection{The Ornstein-Uhlenbeck with Foraging movement model}\label{sec:OUF}
To study the joint effect of range residency and directional persistence on encounter rates, we model individual trajectories using the Ornstein-Uhlenbeck with Foraging (OUF) process \citep{Fleming_et_al_2014}. Recent analyses of large multi-species animal-movement tracking datasets indicate that OUF is the preferred model for the majority of range-resident movement tracks, ahead of the Ornstein-Uhlenbeck (OU) process and independent and identically distributed (IID) movement, particularly when tracking durations are long enough to reveal range residency and sampling resolutions are fine enough to detect directional persistence \citep{Noonan2019,Fagan2025}. OU is the second most selected model, favored when the sampling interval is too coarse to resolve directional persistence, and is recovered as the $\tau_v \rightarrow 0$ limit of OUF, retaining range residency but no velocity autocorrelation. It is also the process underlying earlier analyses of how range residency shapes encounters \citep{Martinez-Garcia_et_al_2020}, which our framework extends by adding the short-time correlations that tracking data resolve.

To maintain mathematical tractability, we will obtain the theoretical results assuming that movement is isotropic and separable in each spatial coordinate. Under these assumptions, the two-dimensional OUF model is defined by a system of stochastic differential equations for the individual location, $\mathbf{z}(t)$, and a correlated noise process, $\mathbf{v}(t)$, which is responsible for generating autocorrelated velocity
\begin{align} 
	\dot{\mathbf{z}}(t)&=-\frac{1}{\tau_{z}} \left[\mathbf{z}(t)-\bm{\lambda}\right]+\mathbf{v}(t),\label{Equation:zdot}\\
	\dot{\mathbf{v}}(t)&=\frac{1}{\tau_{v}} \left[-\mathbf{v}(t)+\sqrt{g}\,\bm{\xi}(t)\right].\label{Equation:vdot}
\end{align}

This system of equations defines a movement pattern characterized by trajectories tending to relax back to a home-range center, $\bm{\lambda}$. This relaxation occurs over a characteristic time scale or home-range crossing time, $\tau_{z}$, thus capturing the typical dynamics of range-resident behavior. The auto-correlated process $\mathbf{v}(t)$ incorporates directional persistence. It is driven by Gaussian white noise $\bm{\xi}(t)$ with zero mean and identity covariance matrix, and introduces a time scale for directional persistence, $\tau_{v}$, which sets the time scale over which velocity correlations decay. This process represents fine-scale foraging or random-search behavior characteristic of many grazing ungulates \citep{Fleming_etal_2014b, Fleming_et_al_2014}. For terrestrial range-resident animals, which are our main focus, $\tau_z \gg \tau_v$ \citep{Morato_et_al_2016, Fleming_et_al_2017, Noonan_et_al_2023}.

The OUF movement model is a Gaussian process because the equations are linear and the only stochastic term, $\bm{\xi}(t)$ in Eq.\,\eqref{Equation:vdot}, is a Normal random variable. The distribution of animal locations thus concentrates probability near the home-range center and decays toward the periphery, reproducing the uneven, center-weighted space use that often typifies range-resident movement \citep{Burt1943}. Additionally, because movement is separable in the $x$ and $y$ coordinates, the two-dimensional process factorizes into two independent one-dimensional processes. Each of them is Gaussian and therefore fully characterized by the first two moments of the location $\mathbf{z}$ and driving process $\mathbf{v}$. Isotropy makes the movement parameters identical in both coordinates, such that the variances coincide and only the means differ through the initial location and the home-range center. We therefore write the moments for a single coordinate, using $z$ and $v$ for the corresponding components of $\mathbf{z}$ and $\mathbf{v}$. The moments of the OUF driving process are
\begin{align}
    \mu_v(t) &= \mu_v(0)\,e^{-t/\tau_v},
    \label{Equation:mu_v(t)}\\[2pt]
    \sigma_v^2(t) &= \sigma_v^2(0)\,e^{-2t/\tau_v}
                     + \frac{g}{2\tau_v}\left(1-e^{-2t/\tau_v}\right),
   \label{Equation:Variance_v(t)}
\end{align}
and the mean location is
\begin{equation}
    \mu_z(t) = \left[\mu_z(0) + \mu_v(0)\,\kappa\left(1-e^{-t/\kappa}\right)\right]
               e^{-t/\tau_z} + \lambda\left(1-e^{-t/\tau_z}\right),
    \label{Equation:mu_z(t)}
\end{equation}
with $\kappa = \tau_z\tau_v/(\tau_z-\tau_v)$ (see Appendix~\ref{app:A} for detailed calculations). Because we integrated the OUF equations from random initial conditions, these moments depend explicitly on the initial values we choose for the mean location $\mu_z(0)$ and the mean and variance of the driving process, $\mu_v(0)$ and $\sigma_v^2(0)$, respectively.
 
The variance of the location thus depends on three initial moments: its own value $\sigma_z^2(0)$, the variance of the driving process $\sigma_v^2(0)$, and the covariance between location and $v$, $\sigma_{zv}(0)$. For most large-bodied terrestrial animals, which are our main interest, the velocity autocorrelation decays over much faster time scales than the location autocorrelation (i.e., $\tau_z \gg \tau_v$) \citep{Noonan2019}. Due to this timescale separation, we sample the initial value of the autocorrelated process $\mathbf{v}$ from its stationary distribution, which is Normal with mean $\mu_v(0)=0$ and variance $\sigma_v^2(0)=g/2\tau_v$. The initial location, in turn, is deterministic, $\sigma_z^2(0)=0$, with $\mu_z(0)$ specifying the actual location of the animal at $t=0$. The Cauchy-Schwarz inequality bounds $\sigma_{zv}(0)$ by
$\sigma_z(0)\sigma_v(0)$ and therefore forces it to vanish as well. Under these conditions, the variance of the location is
\begin{equation}
    \sigma_z^2(t) = \frac{g\,\tau_z}{2\left(\tau_z+\tau_v\right)}
    \left[\tau_z\left(1-e^{-2t/\tau_z}\right)
          - 2\kappa\left(1-e^{-t/\kappa}\right)e^{-2t/\tau_z}\right],
\label{Equation:Variance_z(t)}
\end{equation}
which is a particular case of the more general expression with arbitrary initial conditions (Appendix~\ref{app:A}). In the long-time limit, the moments in Eqs.\,\eqref{Equation:mu_v(t)}-\eqref{Equation:Variance_z(t)} converge to
\begin{align}
    \mu_v(t\rightarrow\infty) = 0, \qquad&
   \sigma_v^2(t\rightarrow\infty) = \frac{g}{2\tau_v}\equiv  S_v,      \label{equation:moments-vel-stationary}
\\[0.6em]
        \mu_z(t\rightarrow\infty) = \lambda, \qquad&
\sigma_z^2(t\rightarrow\infty) = \frac{g\tau_z^2}{2(\tau_z+\tau_v)}\equiv S_z,
    \label{equation:moments-pos-stationary}
\end{align}
where we have introduced $S_v$ and $S_z$ to denote the stationary variance of the driving process $v$ and the location, respectively, because they appear often in the next sections.

\begin{figure}
	\centering	\includegraphics[width=0.85\textwidth]{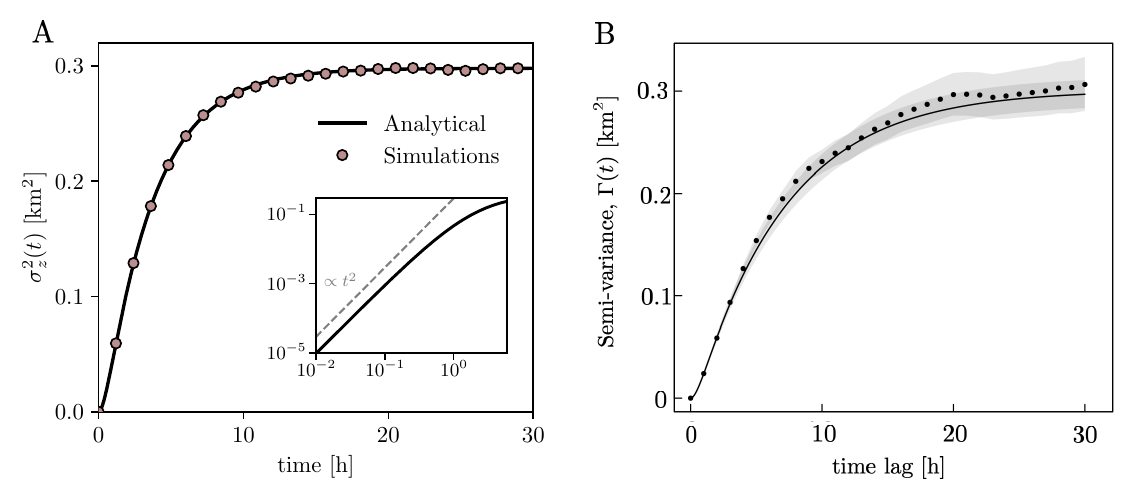}
\caption{A) OUF location variance over time, showing the persistent, BM and OU movement regimes. Lines are analytical, symbols come from $10^6$ simulations of Eqs.\,\eqref{Equation:zdot}-\eqref{Equation:vdot} (Appendix\,\ref{sec:OUFsims}). Initial conditions are deterministic in location, with the driving process drawn from its stationary distribution. B) Empirical variogram (dots) and OUF fit (curve) for a lowland tapir \textit{Tapirus terrestris} trajectory from \cite{Medici_Et_al_2022} (Appendix\,\ref{sec:Variogram_plotting}), with the 95\% confidence interval shaded. Both panels use the parameters returned by this fit,
$\tau_{z} = 6.855$, $\tau_{v} = 0.486$ and
$S_z = 0.298$. Units are kilometers and hours.}
\label{Figure:Variance_OUF_Det_Stoch_ICs}
\end{figure}

Equation\,\eqref{Equation:Variance_z(t)} is a particularly relevant quantity for understanding how OUF movement describes multiple behaviors across different time scales (Fig.\,\ref{Figure:Variance_OUF_Det_Stoch_ICs}A). At short time scales, $\sigma_z^{2}(t)$ increases quadratically with time, reflecting the directional persistence of movement over short intervals (inset in Fig.\,\ref{Figure:Variance_OUF_Det_Stoch_ICs}A). At an intermediate time scale, the variance tends to increase linearly with time, consistent with Brownian motion (BM). This linear relationship suggests that the movement becomes more diffusive, and the direction of motion becomes increasingly uncorrelated over time. Finally, at longer time scales, the location variance approaches an asymptotic value. This plateau sets the limit on the range over which animals can wander, indicating confined or range-resident behavior.

Empirical movement-ecology studies characterize these same time scales through the semi-variance function, or variogram, rather than the variance \citep{Fleming_et_al_2014, Karelus_et_al_2021}. The semi-variance at time lag $t$, $\Gamma(t)$, quantifies the mean squared displacement between locations separated by a time interval $t$,
\begin{equation}
\Gamma(t) = \tfrac{1}{2}\left\langle \left[ \mathbf{z}(t') - \mathbf{z}(t'+t) \right]^2 \right\rangle.
\end{equation}
For a stationary process, the relationship between the semi-variance and the variance of the animal location is 
\begin{equation}
    \Gamma(t) = S_z - C_z(t),
\end{equation}
where $C_z(t)$ is the location autocovariance. Therefore, the variogram grows from zero at short lags and saturates at the stationary variance $S_z$ at long lags. Empirical studies prefer the variogram over the variance because it recovers the same three movement regimes (Fig.\,\ref{Figure:Variance_OUF_Det_Stoch_ICs}B) and it can be estimated from relocation data without knowing the home-range center \citep{Fleming_et_al_2014}. 

\subsection{The OUF mean encounter rate}\label{section:methods-encounter}
To quantify the encounter rates between a pair of OUF individuals, we consider encounters in a dyadic framework \citep{Gurarie2012,Martinez-Garcia_et_al_2020}. This dyadic framework can be readily upscaled to the population level in terms of crowding indexes \citep{Wiegand2021,Menezes_et_al_2025}, and it accurately quantifies direct interactions between organisms, such as mating, competition for space, predation, or disease transmission \citep{Borremans_et_al_2019}. In this dyadic framework, encounters follow a double stochastic process, or reaction-diffusion process in the physics literature \citep{Van_Kampen1992}, such that an encounter may occur only when the separation distance between the two individuals, $r(t)$, is within the encounter distance $q$, which is typically much smaller than each individual's home range.

Conditional on the pair occupying a reactive region $r(t) \le q$, we further assume that encounters follow a Poisson process with constant intensity $\gamma$, with dimensions of area per unit time \citep{Martinez-Garcia_et_al_2020}. This construction separates co-location from a biologically relevant encounter and allows us to explore a continuous range of encounter scenarios, from encounters that occur as soon as two individuals are close enough [strong, or hard,  encounters, $\gamma \rightarrow \infty$ \citep{Gurarie2012,Hutchinson2007}], to encounters that require the two individuals to remain close for long enough on average (weak, or soft, encounters; $\gamma \rightarrow 0$). The probability that an encounter $E$ occurs in the interval $\{t, t+dt\}$ is therefore
\begin{equation}
    \mathbb{P}\left(E\, \mathrm{in}\, \{t, t+dt\}\right) = \mathcal{E}\left(r(t)\right) dt = \gamma\,\Phi_{q}(r(t))\,dt,
    \label{Equation:Prob(E)}
\end{equation}
where the instantaneous encounter rate $\mathcal{E}(r(t)) \equiv \gamma\,\Phi_q(r(t))$ is the probability per unit time of an encounter occurring between a pair of individuals with separation distance $r(t)$ and
\begin{equation}
    \Phi_{q}(r(t)) =
    \begin{cases}
        0 & \text{if } r(t) > q,\\[2pt]
        \dfrac{1}{\pi q^2} & \text{if } r(t) \le q,
    \end{cases}
    \label{Equation:Phi(r(t))}
\end{equation}
defines the space dependence of the reaction rate within the reactive domain. Assuming a uniform value inside the disc and zero outside means that the probability of an encounter happening is independent of the specific distance between individuals, provided this is smaller than the threshold $q$. Additionally, taking this value equal to $1/(\pi q^2)$ normalizes the spatially integrated encounter rate to $\gamma$ independently of $q$, allowing us to interpret $\gamma$ as the total reactivity and $q$ as the threshold in inter-individual distances below which encounters can occur. Because encounters can occur only while $r(t) \le q$, the instantaneous encounter rate is a dichotomous random variable, $\mathcal{E} = \lbrace 0,\, \gamma/\pi q^2 \rbrace$, switching between its two values according to the statistics of pair separation.

To characterize this stochastic process, we define the mean encounter rate as the ensemble average of the instantaneous rate over independent realizations of trajectory pairs. Conditioning on a given initialization of the two velocities, this average reads
\begin{align}
    \bar{\mathcal{E}}\big(t\!\mid\!\mathbf{v}_1(0),\mathbf{v}_2(0)\big) &= \int \mathcal{E}(r)\, f\big(r,\mathbf{v}_1,\mathbf{v}_2,t\!\mid\!r_0,\mathbf{v}_1(0),\mathbf{v}_2(0)\big)\, dr\, d\mathbf{v}_1 d\mathbf{v}_2 \nonumber\\
    &= \frac{\gamma}{\pi q^2} \int_{r \le q} f\big(r,t\!\mid\!r_0,\mathbf{v}_1(0),\mathbf{v}_2(0)\big)\, dr,
    \label{Equation:Mean_Encounter_rate_conditional}
\end{align}
where $f(r,\mathbf{v}_1,\mathbf{v}_2,t\!\mid\!\cdot)$ is the joint density of the pairwise distance and the two velocities at time $t$, and the second equality follows from marginalizing the velocities, because the instantaneous rate depends only on the distance between individuals. Averaging over the initial values of the driving process, which we sample from the stationary distribution $p_{\rm st}$ assumed in the previous section, gives
\begin{equation}
    \bar{\mathcal{E}}(t) = \int \bar{\mathcal{E}}\big(t\!\mid\!\mathbf{v}_1(0),\mathbf{v}_2(0)\big)\, p_{\rm st}\big(\mathbf{v}_1(0)\big)\, p_{\rm st}\big(\mathbf{v}_2(0)\big)\, d\mathbf{v}_1(0)\, d\mathbf{v}_2(0) = \frac{\gamma}{\pi q^2} \int_{r \le q} f\big(r,t\!\mid\!r_0\big)\, dr,
    \label{Equation:Mean_Encounter_rate}
\end{equation}
where $f(r,t\!\mid\!r_0)$ is the density of the pairwise distance after integrating over the initial velocities. Because both initial velocities are sampled from their stationary densities $p_{\rm st}$, their moments are time-independent and $\bar{\mathcal{E}}(t)$ is independent of the initial velocities. For any other initial condition for the driving process $\mathbf{v}$, this dependence decays over $\tau_v$, which is short for most terrestrial mammals \citep{Noonan2019}. Finally, because the OUF process is ergodic, its stationary encounter rate can equivalently be obtained by time-averaging a single sufficiently long trajectory \citep{Figueiredo2025}, making the long-time results of our approach applicable to encounter rates estimated from individual tracking data.

In the limit of local encounters, $q \to 0$, the integral in Eq.~\eqref{Equation:Mean_Encounter_rate} reduces to the density of the pairwise separation vector evaluated at zero \citep{Martinez-Garcia_et_al_2020}. Additionally, because we are assuming that individuals move independently, their joint density factorizes and $f(r,t\!\mid\!r_0)$ follows from the two single-individual location densities,
\begin{equation}
    p_i(\mathbf{z},t) = \int f_i\!\big(\mathbf{z},\mathbf{v},t \!\mid\! \mathbf{z}_i(0)\big)\, d\mathbf{v},
    \label{Equation:position_marginal}
\end{equation}
where the initial condition in each individual's $\mathbf{v}$ has already been averaged over $p_{\rm st}$. The density of the pair separation is the cross-correlation of $p_1$ and $p_2$, so that the mean encounter rate is the inner product of the two individual space-use densities \citep{Martinez-Garcia_et_al_2020,Noonan2021}
\begin{equation}
    \bar{\mathcal{E}}(t) = \gamma \int p_1(\mathbf{z},t)\, p_2(\mathbf{z},t)\, d\mathbf{z}.
    \label{Equation:Mean_encounter_rate_overlap}
\end{equation}
\noindent This expression is valid for any pair of independent movement processes and does not require movement to be Gaussian. Because OUF movement is a Gaussian process, $p_1$ and $p_2$ are Normal at all times, and this overlap integral can be evaluated in closed form, giving
\begin{equation}
\bar{\mathcal{E}}\left(t\right)=\frac{\gamma}{2\pi\sigma_{r}^2(t)}\,
e^{-\frac{\varLambda(t)}{2\sigma_{r}^2(t)}},
\label{Equation:Mean_encounter_rate_qequal0}
\end{equation}
where $\sigma_r^2(t) = \sigma_{z_1}^2(t) + \sigma_{z_2}^2(t)$ is the sum of the variances in the location of the two individuals, and $\Lambda(t) = \mu_{\Delta x}^2(t) + \mu_{\Delta y}^2(t)$ is the squared distance between their mean locations, with $\mu_{\Delta x}(t) = \mu_{x_1}(t) - \mu_{x_2}(t)$ the difference in mean location along each coordinate [see Appendix~\ref{app:B} for a full derivation reproducing \cite{Martinez-Garcia_et_al_2020}].

Both $\sigma_r^2(t)$ and $\Lambda(t)$ are time-dependent, so Eq.\,\eqref{Equation:Mean_encounter_rate_qequal0} describes the encounter rate both during the transient dynamics of the two trajectories and after they have relaxed to the stationary regime. In this stationary regime, the individual mean locations coincide with the home-range centers, and $\Lambda(t)$ equals the squared distance between the centers, $R_\lambda^2$. In this long-time limit, the encounter rate depends on the spatial configuration only through the separation between home-range centers and the home-range sizes, whereas at earlier times it also depends on where, and at what velocity, the individuals started.

\section{Results}

\subsection{The effects of directional persistence depend on home-range overlap} \label{Sec:results_encounter_rate1}

We first investigate how the encounter rate in Eq.~\eqref{Equation:Mean_encounter_rate_qequal0} depends on the time scale for directional persistence, $\tau_v$, and home-range spatial configuration, retaining its full time dependence to characterize both the transient and the stationary regimes.
Throughout all our analyses, we use the initial conditions introduced in Section \ref{sec:OUF}, with the driving process $\mathbf{v}$ sampled from its stationary distribution and a deterministic initial location. In this section and the next, we place individuals at their home-range centers, and we relax that choice in Section \ref{sec:iou}. This initial condition also imposes zero covariance between the location and the driving process by the Cauchy-Schwarz inequality. Finally, we parameterize the movement models with values of the same order as
those obtained for the lowland tapir dataset used in Fig.\,\ref{Figure:Variance_OUF_Det_Stoch_ICs}B. These values are
illustrative rather than a fit to any particular individual. For simplicity, we also assume individuals are identical in their movement parameters and differ only in their home-range spatial location. 

To mimic the conditions of previous studies relating encounter rates to directional persistence \citep{Kiorboe2005,Bartumeus_et_al_2008}, we change the velocity autocorrelation timescale $\tau_v$ while keeping the animal's speed unchanged. Because movement is isotropic and separable in the two coordinates, the stationary mean squared speed is given by
\begin{equation}
    \langle|\dot{\bm z}|^2\rangle = 2S_{\dot{z}} \qquad \text{with} \quad S_{\dot{z}} = \frac{S_z}{\tau_z\tau_v},
\end{equation}
so that $S_z = S_{\dot z}\,\tau_z\,\tau_v$ (see Appendix\,\ref{app:speed} for details on the calculation). Holding the speed fixed as $\tau_v$ varies therefore makes the home-range size grow in direct proportion to the time scale of directional persistence. 

\begin{figure}
	\centering
	\includegraphics[width=\textwidth]{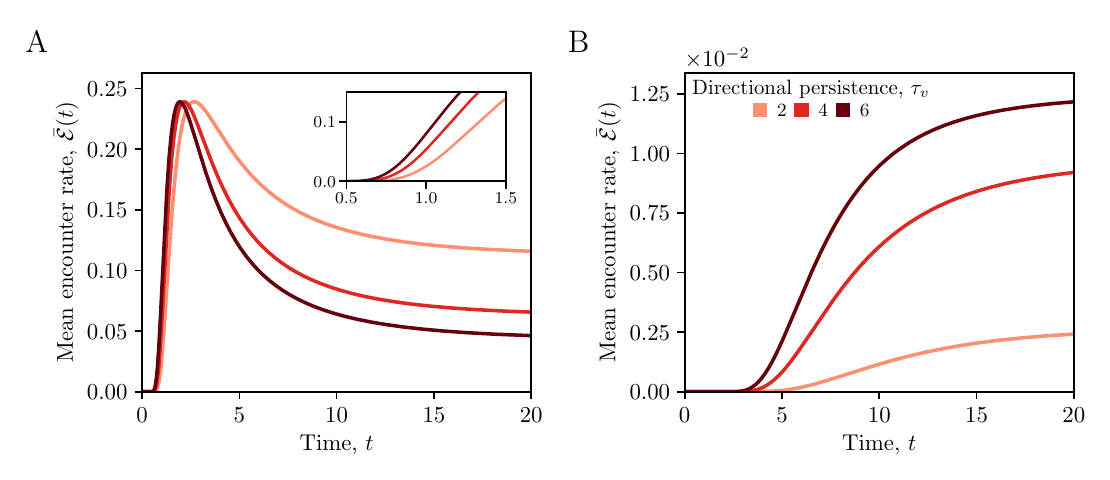}
	\caption{Encounter rate as a function of time for three values of $\tau_v$ in two home-range configurations, $R_\lambda=0.7$ (A) and $R_\lambda=3$ (B). The inset in (A) zooms into the short-time regime. Individuals start at their home-range centers with $\mathbf{v}$ sampled from its stationary distribution. The stationary variance of the velocity $S_{\dot{z}} = 0.0275$ is held constant across the three encounter pairs, ensuring that individuals with different $\tau_v$ move equally fast and differ only in their directional persistence. This requires the driving-process variance $S_v(\tau_v)$ to increase with $\tau_v$. The remaining parameter is $\tau_z=10.341$. Units are kilometers and hours.}
	\label{Figure:OUF_enc_rate}
\end{figure}

We study both a scenario where home-range centers are close to each other and a scenario in which they are farther apart. At short time scales, before range-residency effects become important, persistence over longer time scales enhances encounters, recovering the result obtained for movement without spatial confinement \citep{Bartumeus_et_al_2008, Kiorboe2005}  (Fig.\,\ref{Figure:OUF_enc_rate}). During this transient regime, the encounter rate increases monotonically or peaks at an intermediate time, depending on whether home-range centers are closer or farther apart. More interestingly, when home ranges are sufficiently close, spatial confinement reverses the effect of persistence, such that at long time scales individuals with less persistent movement encounter more often (Fig.\,\ref{Figure:OUF_enc_rate}A). For distant home-range centers and the values of $\tau_v$ we used, more persistent movement leads to higher encounter rates at all times (Fig.~\ref{Figure:OUF_enc_rate}B). Therefore, directional persistence does not have a fixed effect on encounters, and it can either enhance or reduce them depending on the spatial configuration of home ranges.

\subsection{An analytical threshold in range overlap separates the two encounter regimes}\label{sec:threshold}
To better understand what governs this difference, we derive the conditions in home-range size and spatial configuration that distinguish these two regimes, under the same initial conditions used in the previous section. We consider individuals that differ only in the location of their home-range center and have identical movement parameters and home-range size, but generalize our results to the case of different home-range sizes in Appendix \ref{app:bhattacharyya}. When both home ranges are equal in size, the variance of the relative location in each coordinate is $\sigma_r^2 = \sigma_{z_1}^2 + \sigma_{z_2}^2 = 2\sigma_{z}^2$. The mean locations are independent of $\tau_v$ because we initialize $\mathbf{v}$ with zero mean. The encounter rate hence depends on the directional persistence time scale only through $\sigma_z^2(t)$, and using the chain rule we can write
\begin{equation}
    \frac{\partial \bar{\mathcal{E}}}{\partial\tau_v} = \frac{\partial\bar{\mathcal{E}}}{\partial\sigma_z^2}\,\frac{\partial\sigma_z^2}{\partial\tau_v} = \frac{\Lambda - 4\sigma_z^2}{4\sigma_z^4}\,   \bar{\mathcal{E}}\,\frac{\partial\sigma_z^2}{\partial\tau_v},
    \label{Equation:dEdtauv_chain}
\end{equation}
where $\Lambda$ and $\sigma_z^2$, and consequently $\bar{\mathcal{E}}$, are functions of time under the initial conditions we chose. 

Both $\bar{\mathcal{E}}(t)$ and $\sigma_z^2(t)$ are positive, so the sign of \eqref{Equation:dEdtauv_chain} is defined by $\left[\Lambda(t)-4\sigma_z^2(t)\right]$ and $\partial\sigma_z^2/\partial\tau_v$. The second of these two factors is positive over the entire transient (Appendix~\ref{app:dsigma_dtauv}), because at fixed speed a greater $\tau_v$ means that individuals hold their heading for longer and therefore travel farther from their starting point before turning. The sign of $\partial\bar{\mathcal{E}}/\partial\tau_v$ is thus set by the first factor alone. When $\Lambda(t) > 4\sigma_z^2(t)$, the encounters increase with $\tau_v$. In contrast, when $\Lambda(t) < 4\sigma_z^2(t)$, they decrease. The time scale of directional persistence acts only by increasing $\sigma_r^2$, whose effect on the encounter rate is non-monotonic, as already shown for range-resident movement without autocorrelated velocity \citep{Martinez-Garcia_et_al_2020}. Increasing $\tau_v$ can therefore enhance or suppress encounters depending on the spatial organization of the home ranges. Because this argument uses only that $\sigma_r^2$ is a sum of individual variances, each of which increases with $\tau_v$, the criterion is general and valid for individuals differing in other movement parameters.
 
\begin{figure}
	\centering	\includegraphics[width=0.55\textwidth]{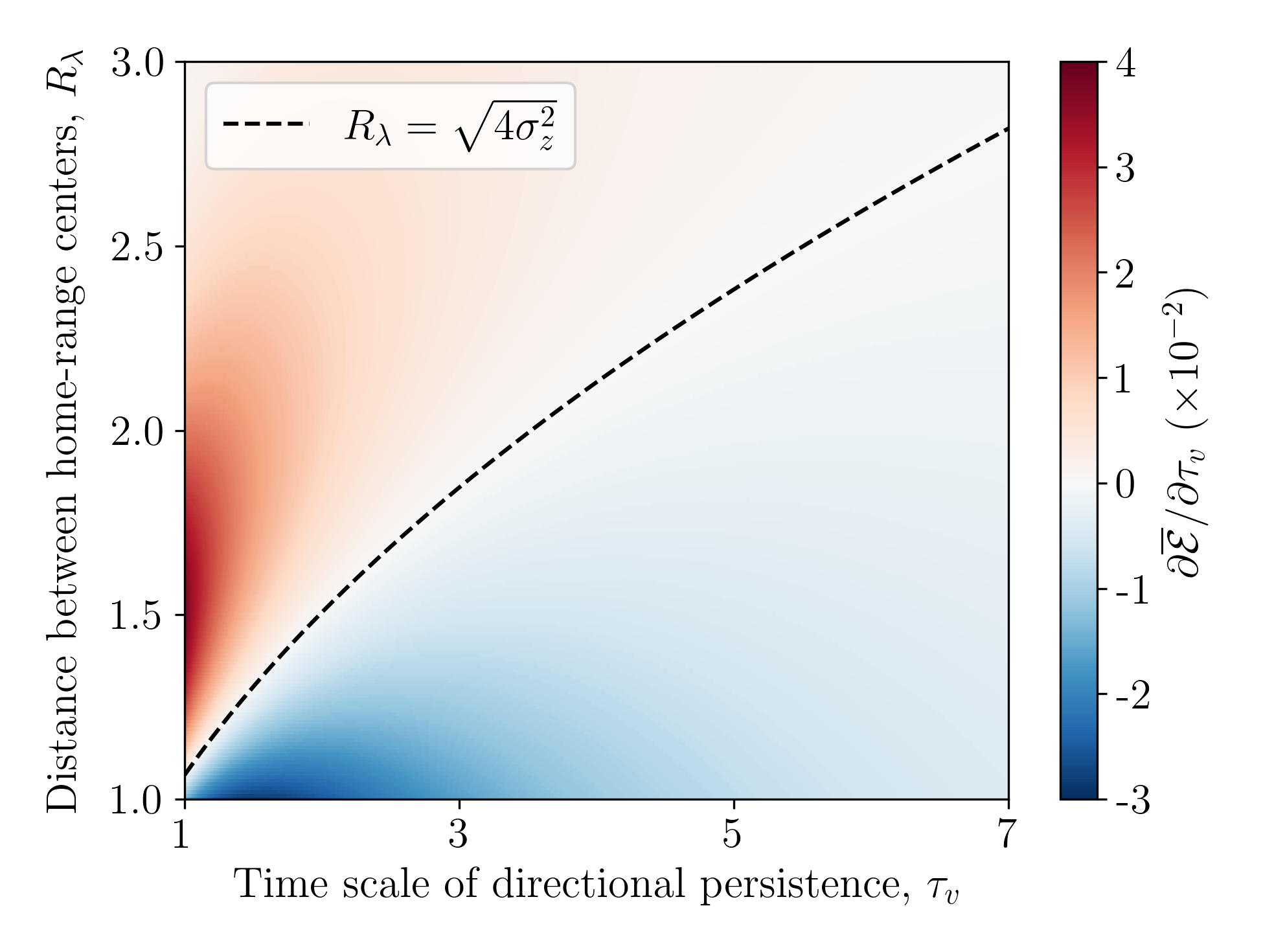}
	\caption{Validation of the analytical threshold in home-range overlap. The color scale shows $\partial\bar{\mathcal{E}}/\partial\tau_v$ over the $(\tau_v, R_\lambda)$ plane and the dashed line is Eq.\,\eqref{Equation:threshold}. Individuals are identical in all movement parameters and differ only in their home-range centers. The stationary variance of the velocity $S_{\dot{z}}$ is kept constant across the plane for both individuals, with $S_v(\tau_v)$ adjusted accordingly to compensate for its dependence on $\tau_v$. For each point, we obtain the mean encounter rate and its derivative with respect to $\tau_v$ by centered differences with step $h=0.05$. Movement parameters are otherwise the same as in Fig.\,\ref{Figure:OUF_enc_rate}, and we sampled $\tau_v$ and $R_\lambda$ with resolutions $0.001$ and $0.01$, respectively.}
	\label{Figure:heatmap-threshold}
\end{figure}

In the long-time limit, after the moments of $\mathbf{z}$ and $\mathbf{v}$ have relaxed to their stationary values, the mean locations coincide with the home-range centers. Consequently, $\Lambda(t\rightarrow\infty)= R_\lambda^2$ and, using Eq.\,\eqref{equation:moments-pos-stationary} for $S_z$ together with $g = 2\tau_v\,S_v$, the threshold condition becomes
\begin{equation}
    R_\lambda = \sqrt{4S_z}
              = 2\sqrt{S_{\dot{z}}\tau_z\tau_v},
    \label{Equation:threshold}
\end{equation}
which separates the two regimes shown in Fig.\,\ref{Figure:OUF_enc_rate} and reproduces the sign change of $\partial\bar{\mathcal{E}}/\partial\tau_v$ across the whole $(\tau_v, R_\lambda)$ plane, covering $\tau_v/\tau_z$ ratios higher than those typically estimated for terrestrial mammals (Fig.\,\ref{Figure:heatmap-threshold}). At fixed speed, more persistent movement enlarges the home range in direct proportion to $\tau_v$, so the threshold grows as $\sqrt{\tau_v}$ and individuals must be farther apart for persistence to increase their encounter rate. Inverting Eq.\,\eqref{Equation:threshold} gives, for a pair of organisms with home-range centers separated by a distance $R_\lambda$, the threshold in the time scale of directional persistence beyond which more persistent movement starts to reduce the encounter rate,
\begin{equation}
    \tau_v^\ast = \frac{R_\lambda^2}{4\,S_{\dot z}\,\tau_z},
\end{equation}
showing that every spatial configuration reaches this reversal at a finite value of $\tau_v$.
However, because parameterizations from tracking data are constrained to $\tau_z \gg \tau_v$, only pairs with sufficiently overlapping home ranges reach the reversal within the observed range of directional persistence.

The condition in Eq.\,\eqref{Equation:threshold} compares the width of the two location densities with the separation between their means, and both quantities are time dependent. It is thus natural to rewrite it in terms of home-range overlap metrics that are more common in spatial ecology \citep{Winner_et_al_2018} and that have already been used to express the encounter rate itself \citep{Martinez-Garcia_et_al_2020}. Here we use the Bhattacharyya distance, which measures the distance between two probability distribution functions and is defined as
\begin{equation}
D_B = -\ln \int d\mathbf{z}\,\sqrt{p_1(\mathbf{z})\,p_2(\mathbf{z})},
    \label{Equation:bhattacharyya}
\end{equation}
where $p_i(\mathbf{z})$ are the space utilization functions for each individual \citep{Fleming2025}. For the simple case we are considering here of animals moving according to isotropic OUF models, $D_B$ can be straightforwardly computed, allowing us to rewrite the threshold in Eq.\,\eqref{Equation:threshold} as $D_B = 1/2$. Therefore, for animals with identical movement time scales and home-range size, increasing directional persistence favors encounters for $D^*_B > 1/2$ and diminishes them otherwise. When the two home ranges differ in size, the transition still occurs at $R_\lambda^2 = 4\bar{S}_z$, with $\bar{S}_z$ the mean of the two home-range variances, but the corresponding critical Bhattacharyya distance is displaced to
\begin{equation}
D_B^{*} = 1/2 + \ln\left(\frac{\bar{S}_z}{\sqrt{S_{z_1}S_{z_2}}}\right) > 1/2
\end{equation}
(see Appendix~\ref{app:bhattacharyya} for both calculations). This result shows that even if persistence acts over short time scales, whether it favors encounters or not depends on home-range spatial arrangement, which is a long-term movement property.

\subsection{Encounter statistics diverge earlier than movement statistics}
\label{sec:iou}

Previous sections considered individuals starting at their home-range centers. In this scenario, their mean locations and $\Lambda$ remain constant, and the effect of directional persistence enters only through $\sigma_r^2$. Because the encounter rate depends on both $\sigma_r^2$ and $\Lambda$, we now consider scenarios where individuals are initially away from their centers, such that $\Lambda$ changes over time as their mean locations relax to the home-range centers. 

\begin{figure}
	\centering	\includegraphics[width=\textwidth]{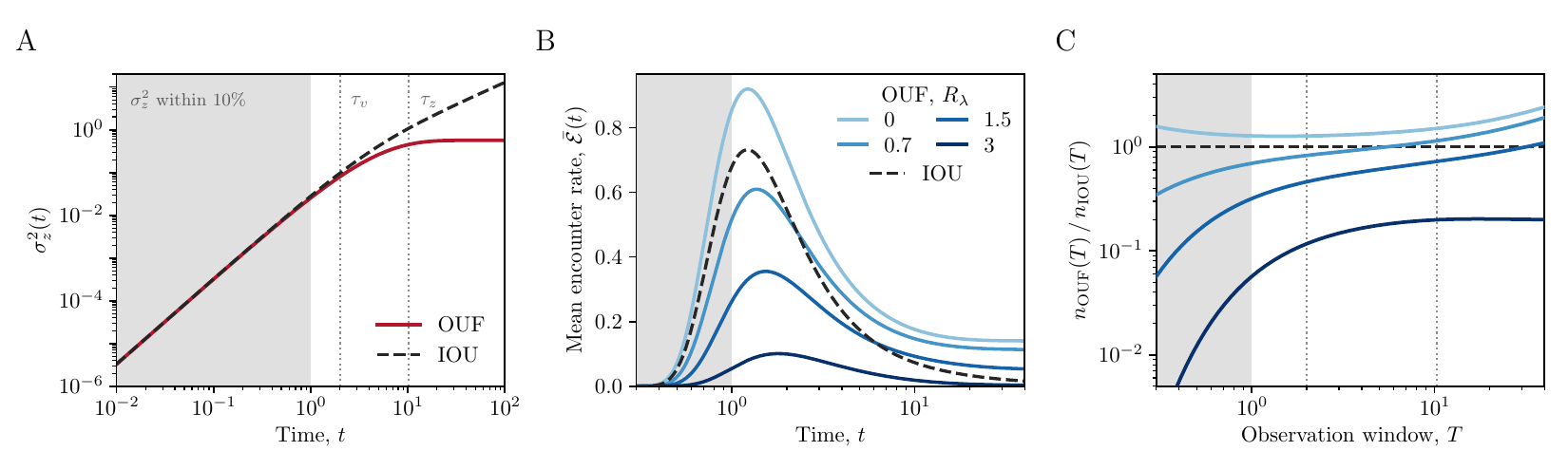}
	\caption{Neglecting range residency changes the encounter statistics before it changes the movement statistics (i.e., sooner after release). The two individuals are released a distance $d_0=0.4$ apart on the line through their home-range centers, which are separated by $R_\lambda$, with velocities sampled from the stationary distribution. IOU is the $\tau_z\rightarrow\infty$ limit of OUF, with the same directional persistence and release speed but no home-range center, so it gives a single prediction for every $R_\lambda$. A) Variance of the animal location for OUF, Eq.\,\eqref{Equation:Variance_z(t)}, and for IOU, Eq.\,\eqref{Equation:sigma_iou}. The shaded region marks where the two agree to within $10\%$. B) Mean encounter rate, Eq.\,\eqref{Equation:Mean_encounter_rate_qequal0}, which has already separated inside that region. C) Expected number of encounters over a window $T$, Eq.\,\eqref{Equation:n_encounters}, relative to the IOU prediction. Remaining parameters are $\tau_v=2$, $\tau_z=10.341$ and $S_{\dot{z}}=0.0275$. Units are kilometers and hours.} \label{Figure:IOU_vs_OUF}
\end{figure}

To isolate the effect of this relaxation, we compare the resulting encounters with those predicted by a movement model without range residency, in which the variance of the animal location grows in the same way at short times but the mean locations remain fixed. Taking the limit $\tau_z\rightarrow\infty$ in the OUF equations, Eqs.\,\eqref{Equation:zdot}-\eqref{Equation:vdot}, gives such a model, the integrated Ornstein-Uhlenbeck process (IOU), which preserves OUF directional persistence but lacks long-term spatial confinement (Fig.\,\ref{Figure:IOU_vs_OUF}A). In this limit, the driving process $\mathbf{v}$ gives the animal velocity, and the variance of the animal location in Eq.\,\eqref{Equation:Variance_z(t)} becomes
\begin{equation}
    \sigma_{z,\mathrm{IOU}}^2(t) = 2\,S_v\,\tau_v
    \left[t - \tau_v\left(1 - e^{-t/\tau_v}\right)\right],
    \label{Equation:sigma_iou}
\end{equation}
which behaves ballistically at short times $\sigma_z^2(t)\simeq S_vt^2$ (like the OUF process), but grows unbounded at long times with an effective diffusion coefficient $S_v\,\tau_v$ that increases with the time scale of the velocity autocorrelation (Fig.\,\ref{Figure:IOU_vs_OUF}A). This is also the model that analyses of tracking data select when the observation window is too short for range residency to become apparent \citep{Noonan2019}.

To compare the two models, we consider a scenario in which a pair of individuals are initially located a distance $d_0$ apart on the line through their home-range centers, which are separated by $R_\lambda$. We chose $d_0$ and $R_\lambda$ to ensure that both animals always start within their home range (i.e., the $95$th percentile of their space utilization function). For simplicity, we assume that both home-range centers and initial animal locations are on the $x$ axis, so individual initial locations are $(\pm\, d_0/2,0)$ and the home-range centers are at $(\pm R_\lambda/2,0)$. With these initial conditions, the squared distance between the mean animal locations is
\begin{equation}
\Lambda(t) = \left[d_0 e^{-t/\tau_z} + R_\lambda\left(1 - e^{-t/\tau_z}\right)\right]^2
\end{equation}
for OUF movement. Because the IOU model does not account for range residency, the mean animal locations do not move and $\Lambda(t) = d_0^2$ at all times. The velocities are sampled from the stationary distribution as in the previous sections.

The encounter rates predicted by the two models diverge quickly at short times (Fig.\,\ref{Figure:IOU_vs_OUF}B), even while the variances of the animal locations remain close to each other (shaded region in Fig.\,\ref{Figure:IOU_vs_OUF}A). This divergence appears because the encounter rate depends on both the variance in the animal locations and their mean locations. For IOU movement, the mean locations stay where the individuals started. However, for OUF, they relax towards the home-range centers, and $\bar{\mathcal{E}}$ depends on their separation exponentially through $\Lambda$. When the home-range centers are closer together than the starting locations, the mean locations converge, and encounters become more frequent than IOU predicts. Conversely, when the home-range centers are farther apart than the starting locations, the mean locations separate and encounters become less frequent.

Finally, we obtain a more direct comparison between movement models and home-range spatial configuration by measuring how encounters accumulate over a given observation time. The expected number of encounters within an observation time $T$ is given by
\begin{equation}
    n(T) = \int_0^T dt\;\bar{\mathcal{E}}(t),
    \label{Equation:n_encounters}
\end{equation}
which we can compute using Eq.\,\eqref{Equation:Mean_encounter_rate_qequal0} with model-specific value of $\Lambda(t)$ introduced above. The accumulated number of encounters varies widely across spatial configurations, falling both above and below the single value predicted by IOU even at short times when OUF and IOU variances still agree (Fig.\,\ref{Figure:IOU_vs_OUF}C). Therefore, over observation windows too short for range residency to become apparent, the choice of movement model has little effect on how well the trajectories are described but a large effect on the encounter predictions derived from them.

\section{Discussion}

We derived exact expressions for the encounter rate between two individuals whose movement combines short-time autocorrelated velocity with long-time range residency, and used them to establish how directional persistence shapes encounters when animals confine their movement to a home range. At constant speed, more persistent movement enlarges the home range in direct proportion to the time scale of directional persistence, which lowers the encounter rate between individuals with strong home-range overlap and raises it when overlap is low. We derived the threshold at which this sign changes in terms of the Bhattacharyya distance, which can be routinely estimated from tracking data \citep{Winner_et_al_2018}. Comparing these results with those obtained for movement without range residency, we found that neglecting range residency leads to inaccurate estimates of encounter frequency, even over time scales too short for confinement to be visible in the variance of the animal location (Fig.\,\ref{Figure:IOU_vs_OUF}). This occurs because range residency not only bounds the growth of the location variance, but also displaces the mean animal locations unless individuals start exactly at their home-range centers.

At short times, the variance in the animal location grows from zero and is still small compared with the separation between home-range centers, so more persistent movement increases encounter rates for any arrangement of the two home ranges. This is the same result obtained in previous studies using unconfined movement models, and it requires that individuals start at a known location, as assumed in those studies \citep{Bartumeus_et_al_2008,Kiorboe2005,Visser_and_Kiorboe_2006}. Those studies are restricted to this short-time regime because their movement models become Brownian at long time scales, whereas for many species real trajectories remain confined to a home range \citep{Bartumeus_et_al_2008}. Using movement models that capture this confinement, we find that once range residency saturates the variance in each animal's location, higher directional persistence continues to increase encounter frequency between pairs with weakly overlapping ranges but reduces encounter frequency between pairs with strongly overlapping ranges. The long-term effect of directional persistence on encounters is therefore not universal and depends on the spatial organization of home ranges.

Encounters depend on both the mean animal locations and the variance around them. When animals start at their home-range centers, their mean locations stay fixed and only the variance changes, so range residency affects encounters only at long times, once the variance saturates \citep{Fleming_et_al_2014,Martinez-Garcia_et_al_2020}. When they start anywhere else, their mean locations begin relaxing toward the centers immediately, and encounters change from the outset, long before the variance saturates. This asymmetry matters because deciding whether a track is range resident rests on its variogram, as in the \textit{ctmm} workflow \citep{Calabrese_et_al_2016}. Variograms are built from the variance alone and plateau only once the sampling duration exceeds the home-range crossing time \citep{Fleming_et_al_2014,Silva_et_al_2022}. Shorter tracks must therefore be fitted with non-resident models, in which the mean locations never move, and only longer sampling can remove this limitation. Resolving both movement behaviors also constrains study design, because sampling is usually optimized either for long duration, which reveals range residency, or for high frequency, which reveals directional persistence \citep{silva2026too}. Encounters require both, which is demanding for species whose tracking devices cannot sustain fine sampling over long deployments.

Range residency thus shapes encounters at time scales shorter than those at which it can be identified in animal tracks, so that the choice of movement model has little effect on how well short trajectories are described but a large effect on the encounters predicted from them. These errors can be large (Fig.\,\ref{Figure:IOU_vs_OUF}C), so deviations from mass-action predictions \citep{Hutchinson2007,Figueiredo2025} will be especially strong for particular home-range arrangements and observation windows. These misestimates of encounter rates would, in turn, have considerable consequences for ecological interactions that hinge on encounters, such as mating, predation, and disease transmission \citep{Turesson_and_Brönmark_2007,Wilber_et_al_2022}.

Once tracks are long enough to resolve range residency, the effect of the time scale of directional persistence can be measured directly using estimators of the Bhattacharyya distance \citep{Fieberg2005,Winner_et_al_2018}. These estimates also give overlap metrics the encounter-relevant spatial scale they otherwise lack, providing a basis for studying the causes of home-range spatial configurations in wild populations \citep{arbon_dwarf_2026,Fagan2024,Potts2014,Torrez2020}. Because the threshold depends on the separation between home-range centers, processes that pack individuals closer together, such as habitat loss, move pairs to the side where more directionally persistent movement lowers rather than raises the frequency of encounters. The consequences of crowding for encounters therefore depend on how animals move, and not on density alone.

This approach can also be generalized to investigate interactions involving multiple organisms \citep{Menezes_et_al_2025}. Specifically, our theoretical finding that increasing directional persistence can enhance or diminish encounter rates depending on home-range overlap suggests that animals could adapt their movement persistence to regulate encounters with conspecifics and heterospecifics. In fact, telemetry data on polar bears showed that adult females tend to exhibit more directionally persistent movement to optimize prey availability, whereas adult males exhibit more tortuous movement to reduce male-male encounters \citep{Laidre_2013}. Similar adaptive trends have recently been observed across prey and predator terrestrial mammals, indicating that animals change movement persistence depending on resource availability and the selective pressures acting on their movement strategies \citep{Noonan_et_al_2023}. Animals also change their movement over time, shifting their activity toward hours with less predator activity \citep{moretto_free-ranging_2026} or altering their space use upon infection \citep{tao_neglecting_2026}. Future work should thus incorporate time dependence in movement behaviors.

In summary, we developed an analytically tractable framework that connects important multiscale movement behaviors to the encounter rates that drive population-level processes. Tracking devices now provide high-resolution movement data \citep{Nathan2022}, and new statistical methods can analyze the highly autocorrelated trajectories they generate \citep{Silva_et_al_2022}, giving an unprecedented understanding of how animals move \citep{Fleming_et_al_2014,Vilk2022}. These developments have outpaced ecological theory, particularly regarding encounter and interaction models. Our study provides a first step toward closing this gap, showing that the multiscale structure of movement can reverse long-standing expectations about how movement shapes encounters, and that encounter rates are a more demanding target for inference than the movement parameters themselves.

\section*{Acknowledgements}
This work was partially funded by the Center for Advanced Systems Understanding (CASUS), which is financed by Germany’s Federal Ministry of Research, Technology and Space (BMFTR) and by the Saxon Ministry for Science, Culture and Tourism (SMWK) with tax funds on the basis of the budget approved by the Saxon State Parliament. RMG received partial support from FAPESP through ICTP-SAIFR 2021/14335-0. WFF was supported by US NSF award DMS2451241.

\bibliography{references}

\vspace{1.5cm}

\noindent \textbf{\Large Supporting Information}

\begin{appendices}

\titleformat{\section}[display]{\normalfont\Large\bfseries}{\appendixname~\thesection.}{0.4ex}{}

\section{Calculation of the moments for the OUF model}\label{app:A}

To compute the moments of the OUF model, we first integrate the coupled stochastic differential equations of the OUF movement model, given by Eq.\,\eqref{Equation:zdot} and Eq.\,\eqref{Equation:vdot} in the main text. The solutions, obtained through direct integration, are
\begin{align}
    z(t) &= z(0)\,e^{-t/\tau_z} + \lambda\left(1-e^{-t/\tau_z}\right)
            + \int_0^t ds\; v(s)\,e^{-(t-s)/\tau_z},
            \label{Equation:z_solution-SM}\\[2pt]
    v(t) &= v(0)\,e^{-t/\tau_v}
            + \frac{\sqrt{g}}{\tau_v}\int_0^t dW\, e^{-(t-s)/\tau_v},
            \label{Equation:v_solution-SM}
\end{align}
where $dW = \xi(s)\,ds$. Because the system of SDEs is linear and the noise enters the location equation only through $v(s)$, both $v$ and $z$ are linear functionals of a Gaussian white noise. The joint process $(z,v)$ is therefore Gaussian whenever the initial conditions are Gaussian or deterministic, and is completely specified by the moments computed below. The linearity of the equations also implies that the location is a deterministic linear functional of $\mathbf{v}$, so we can obtain every moment of $\mathbf{z}$ by ordinary integration once the moments of $\mathbf{v}$ are known.\\
 
\noindent{\textit{Mean values}---}We first obtain the mean of each component $\mathbf{v}$ by averaging Eq.\,\eqref{Equation:v_solution-SM} and using $\langle dW\rangle = 0$,
\begin{equation}
    \mu_v(t) = \mu_v(0)\,e^{-t/\tau_v},
    \label{Equation:mu_v-SM}
\end{equation}
where $\mu_v(0)$ is the mean value initial value of $\mathbf{v}$. Next, averaging Eq.\,\eqref{Equation:z_solution-SM} and inserting Eq.\,\eqref{Equation:mu_v-SM} gives the mean location,
\begin{equation}
    \mu_z(t) = \mu_z(0)\,e^{-t/\tau_z} + \lambda\left(1-e^{-t/\tau_z}\right)
               + \mu_v(0)\,A(t),
    \label{Equation:mu_z-SM}
\end{equation}
where $\mu_z(0)$ is the mean initial location and we have defined the response
function
\begin{equation}
    A(t) \equiv \int_0^t ds\; e^{-s/\tau_v}\,e^{-(t-s)/\tau_z}
         = \kappa\left(1-e^{-t/\kappa}\right)e^{-t/\tau_z},
    \label{Equation:response_function}
\end{equation}
and $\kappa = \tau_z\tau_v/(\tau_z-\tau_v)$ as defined in the main text. $A(t)$ measures how a fluctuation in $\mathbf{v}$ propagates into the animal location after a time $t$, vanishing at $t=0$ and at $t\rightarrow\infty$.\\
 
\noindent{\textit{Variances---}}To obtain the variance of the $\mathbf{v}$, we first compute its two-time covariance. Defining $\delta v(t) = v(t) - \mu_v(t)$ and using Eq.\,\eqref{Equation:v_solution-SM}, the two time covariance is
\begin{equation}
    C_v(t_2,t_1) = \left\langle \delta v(t_2)\,\delta v(t_1)\right\rangle
    = \sigma_v^2(0)\,e^{-(t_2+t_1)/\tau_v}
      + \frac{g}{\tau_v^2}\left\langle \int_0^{t_2}\!\!\int_0^{t_1}
        dW_s\,dW_u\; e^{-(t_2+t_1-s-u)/\tau_v}\right\rangle ,
    \label{Equation:Cv_raw}
\end{equation}
where $\sigma_v^2(0)$ is the initial variance of the autocorrelated process $\mathbf{v}$. Using
$\langle dW_s\, dW_u\rangle = \delta(s-u)$, the double integral becomes a single integral taken over the overlap of the two integration time windows, $0$ to $\mathrm{min}(t_1,t_2)$. Using $t_2+t_1-2\min(t_2,t_1) = |t_2-t_1|$, we get
\begin{equation}
    C_v(t_2,t_1) = \sigma_v^2(0)\,e^{-(t_2+t_1)/\tau_v}
    + \frac{g}{2\tau_v}\left[e^{-|t_2-t_1|/\tau_v} - e^{-(t_2+t_1)/\tau_v}\right].
    \label{Equation:Cv_two_time}
\end{equation}
Finally, considering the particular case $t_2=t_1=t$, we obtain the variance of one component of $\mathbf{v}$
\begin{equation}
    \sigma_v^2(t) = \sigma_v^2(0)\,e^{-2t/\tau_v}
                    + \frac{g}{2\tau_v}\left(1-e^{-2t/\tau_v}\right),
    \label{Equation:sigma_v}
\end{equation}
where the first term on the right side contains the decay of the initial value of the variance and the second one is the contribution built up by the noise that vanishes at $t=0$.
 
We next work on the location variance. Subtracting Eq.\,\eqref{Equation:mu_z-SM} from
Eq.\,\eqref{Equation:z_solution-SM}, every deterministic contribution cancels,
including the terms in $\lambda$, and what remains involves only the deviations
of $z(0)$ and $v(s)$ from their means,
\begin{equation}
    \delta z(t) = \delta z(0)\,e^{-t/\tau_z}
                  + \int_0^t ds\;\delta v(s)\,e^{-(t-s)/\tau_z}.
    \label{Equation:z_fluctuation}
\end{equation}

Multiplying two copies of Eq.\,\eqref{Equation:z_fluctuation} evaluated at $t_2$
and $t_1$ and averaging, the exponentials are deterministic and come outside, so
the average acts only on products of $\delta z(0)$ and $\delta v$,
\begin{align}
    C_z(t_2,t_1) = \;& \sigma_z^2(0)\,e^{-(t_2+t_1)/\tau_z}
      + e^{-t_2/\tau_z}\!\int_0^{t_1}\!\! ds'\, C_{zv}(0,s')\,e^{-(t_1-s')/\tau_z}
      + e^{-t_1/\tau_z}\!\int_0^{t_2}\!\! ds\, C_{zv}(0,s)\,e^{-(t_2-s)/\tau_z}
      \nonumber\\[2pt]
      &+ \int_0^{t_2}\!\!\int_0^{t_1} ds\,ds'\;
         e^{-(t_2-s)/\tau_z}\,e^{-(t_1-s')/\tau_z}\;C_v(s,s'),
    \label{Equation:Cz_from_Cv}
\end{align}
where $\sigma_z^2(0)$ is the variance in initial location and the  covariance between the animal location and the OUF driving process $C_{zv}(0,s) = \left\langle \delta z(0)\,\delta v(s)\right\rangle
= \sigma_{zv}(0)\,e^{-s/\tau_v}$ follows from Eq.\,\eqref{Equation:v_solution-SM}, because the noise driving the $\mathbf{v}$ at $t>0$ is independent of the initial state.
 
The first term of Eq.\,\eqref{Equation:Cz_from_Cv} is already explicit, so to obtain a closed expression for the location variance we set $t_2=t_1=t$ and evaluate the remaining integrals. The two cross terms become identical and, inserting $C_{zv}(0,s) = \sigma_{zv}(0)\,e^{-s/\tau_v}$ using the definition of $A(t)$ in Eq.\,\eqref{Equation:response_function}, we get
\begin{equation}
    e^{-t/\tau_z}\!\int_0^{t}\!\! ds\;\sigma_{zv}(0)\,e^{-s/\tau_v}\,
    e^{-(t-s)/\tau_z}
    = \sigma_{zv}(0)\,e^{-t/\tau_z}\!\!
      \int_0^{t}\!\! ds\; e^{-s/\tau_v}e^{-(t-s)/\tau_z}
    = \sigma_{zv}(0)\,e^{-t/\tau_z}A(t),
    \label{Equation:cross_term}
\end{equation}
for each of them. This term is non-zero only when the initial value of the animal location and the OUF driving process are correlated, and vanishes for $\sigma_{zv}(0)=0$.

The final term we need to compute is the double integral of Eq.\,\eqref{Equation:Cz_from_Cv},
\begin{equation}
    D(t) \equiv \int_0^{t}\!\!\int_0^{t} ds\,ds'\;
    e^{-(t-s)/\tau_z}\,e^{-(t-s')/\tau_z}\;C_v(s,s').
    \label{Equation:D_definition}
\end{equation}
Before integrating, we regroup Eq.\,\eqref{Equation:Cv_two_time} by collecting the two terms that share the same exponential,
\begin{equation}
    C_v(s,s') = \left[\sigma_v^2(0)-\frac{g}{2\tau_v}\right]
                 e^{-s/\tau_v}\,e^{-s'/\tau_v}
              + \frac{g}{2\tau_v}\,e^{-|s-s'|/\tau_v}.
    \label{Equation:Cv_regrouped}
\end{equation}
This rewriting of $C_v(s,s')$ separates the decay of the initial condition, weighted by the difference between $\sigma_v^2(0)$ and its stationary value $g/2\tau_v$, from the stationary autocorrelation built up by the noise, which depends on the two times only through their difference. In the first contribution, the integrand factorizes into a function of $s$ times a function of $s'$. The double integral then becomes the product of two identical single integrals, which, using the definition of $A(t)$ in Eq.\,\eqref{Equation:response_function} can be written as
\begin{equation}
    \int_0^{t}\!\!\int_0^{t} ds\,ds'\;
    e^{-(t-s)/\tau_z}e^{-(t-s')/\tau_z}\,e^{-s/\tau_v}e^{-s'/\tau_v}
    = \left[\int_0^{t}\!\! ds\; e^{-s/\tau_v}e^{-(t-s)/\tau_z}\right]^{2}
    = A^2(t).
    \label{Equation:I1_result}
\end{equation}

The second contribution does not factorize, because $e^{-|s-s'|/\tau_v}$ couples the two integration variables. We write this term as
\begin{equation}
    N(t) \equiv \int_0^{t}\!\!\int_0^{t} ds\,ds'\;
    e^{-(t-s)/\tau_z}\,e^{-(t-s')/\tau_z}\;e^{-|s-s'|/\tau_v},
    \label{Equation:N_definition}
\end{equation}
such that
\begin{equation}
    D(t) = \left[\sigma_v^2(0)-\frac{g}{2\tau_v}\right]A^2(t)
+ \frac{g}{2\tau_v}N(t).
\end{equation}
 
To evaluate $N(t)$, we use the symmetry of its integrand under $s\leftrightarrow s'$, which allows us to restrict the integration to the half-domain $s>s'$, where $|s-s'| = s-s'$ just by multiplying the whole integral by a factor two,
\begin{equation}
    N(t) = 2\,e^{-2t/\tau_z}\int_0^{t}\!\! ds\;
           e^{s/\tau_z}e^{-s/\tau_v}
           \int_0^{s}\!\! ds'\; e^{s'/\tau_z}e^{s'/\tau_v}.
    \label{Equation:N_halfdomain}
\end{equation}
We next introduce the time scale
\begin{equation}
    \bar{\tau} \equiv \frac{\tau_z\tau_v}{\tau_z+\tau_v},
\end{equation}
which together with $\kappa$ satisfy $1/\bar{\tau}-1/\kappa = 2/\tau_z$. With these definitions the inner integral of Eq.\,\eqref{Equation:N_halfdomain} is
\begin{equation}
\int_0^{s} ds'\,e^{s'/\bar{\tau}} = \bar{\tau}(e^{s/\bar{\tau}}-1),
\end{equation}
while the prefactor of the outer integrand is $e^{s/\tau_z}e^{-s/\tau_v} = e^{ s/\kappa}$,
so that
\begin{align}
    N(t) &= 2\bar{\tau}\,e^{-2t/\tau_z}\left[
            \int_0^{t}\!\! ds\; e^{s\left(1/\bar{\tau}-1/\kappa\right)}
          - \int_0^{t}\!\! ds\; e^{-s/\kappa}\right]
          \nonumber\\[2pt]
         &= 2\bar{\tau}\,e^{-2t/\tau_z}\left[
            \frac{\tau_z}{2}\left(e^{2t/\tau_z}-1\right)
          - \kappa\left(1-e^{-t/\kappa}\right)\right],
    \label{Equation:N_intermediate}
\end{align}
Using once again the definition of $A(t)$ in Eq.\,\eqref{Equation:response_function}, we obtain the closed form
\begin{equation}
    N(t) = \bar{\tau}\,\tau_z\left(1-e^{-2t/\tau_z}\right)
         - 2\,\bar{\tau}\,e^{-t/\tau_z}A(t),
    \label{Equation:N_result}
\end{equation}
where the first term is the variance that a purely diffusive process would accumulate under the confinement set by $\tau_z$, and the second is the correction due to the finite correlation time in $\mathbf{v}$. This last term and $A(t)$ vanish both at $t=0$ and as $t\rightarrow\infty$.
 
Collecting all terms in Eqs.\,\eqref{Equation:cross_term}, \eqref{Equation:I1_result} and \eqref{Equation:N_result}, the variance of the location can be written as
\begin{equation}
    \sigma_z^2(t) = \sigma_z^2(0)\,e^{-2t/\tau_z}
                  + 2\,\sigma_{zv}(0)\,e^{-t/\tau_z}A(t)
                  + \left[\sigma_v^2(0)-\frac{g}{2\tau_v}\right]A^2(t)
                  + \frac{g}{2\tau_v}\,N(t),
    \label{Equation:sigma_z}
\end{equation}
where the first term is the decay of the initial location variance, the second transfers the initial correlation between the animal location and the OUF driving process into the location, the third propagates the excess initial driving-process variance through $A(t)$, and the last one is the variance built up by the noise. At $t=0$ both $A(t)$ and $N(t)$ vanish, recovering $\sigma_z^2(0)$. On the other hand, when $t\rightarrow\infty$, every term containing $e^{-t/\tau_z}$ or $A(t)$ decays to $0$ and $N(t\to \infty)=\bar{\tau}\tau_z$. Consequently, the stationary variance becomes
\begin{equation}
    \sigma_z^2(\infty) = \frac{g\,\tau_z^2}{2\left(\tau_z+\tau_v\right)},
    \label{Equation:sigma_z_stationary}
\end{equation}
which, as expected, is independent of the initial condition.

Finally, if we particularize Eq.\,\eqref{Equation:sigma_z} for the initial conditions used in the main text (deterministic location and driving process sampled from the stationary distribution), we recover Eq.\,\eqref{Equation:Variance_z(t)}. Under these initial conditions, the first term in Eq.\,\eqref{Equation:sigma_z} vanishes because $\sigma_z^2(0)=0$. The second term, because the Cauchy-Schwarz inequality imposes $\sigma_{zv}(0)=0$ if $\sigma_z^2(0)=0$. Finally, the third term vanishes because imposing that $\mathbf{v}$ is initially at stationarity implies $\sigma_v^2(0)-g/2\tau_v=0$. Inserting Eq.\,\eqref{Equation:N_result} and using $\bar{\tau}=\tau_z\tau_v/(\tau_z+\tau_v)$ together with $A(t)=\kappa\left(1-e^{-t/\kappa}\right)e^{-t/\tau_z}$, the factors $1/\tau_v$ cancel, and we obtain
\begin{equation}
    \sigma_z^2(t) = \frac{g\,\tau_z}{2\left(\tau_z+\tau_v\right)}
    \left[\tau_z\left(1-e^{-2t/\tau_z}\right)
          - 2\kappa\left(1-e^{-t/\kappa}\right)e^{-2t/\tau_z}\right],
    \label{Equation:sigma_z_ic_compact}
\end{equation}
which corresponds to Eq.\,\eqref{Equation:Variance_z(t)} in the main text.


\section{Numerical integration of the OUF stochastic differential equations}\label{sec:OUFsims}
The system of stochastic differential equations describing the OUF movement, Eq.\,\eqref{Equation:zdot}-\eqref{Equation:vdot} is numerically integrated using the Euler-Maruyama method over a uniform time grid of step size $\Delta t=0.01\,\textrm{hr}$. The Euler-Maruyama method evaluates the deterministic part of the SDE explicitly, while approximating the stochastic terms with a normal random increment with mean zero and variance $\Delta t$. If we write the location and driving process at $n^{\textrm{th}}$ time step as $\mathbf{z}_n=\mathbf{z}(n\Delta t)$ and $\mathbf{v}_n=\mathbf{v}(n\Delta t)$, then the update rules are given by,
\begin{align}
	\mathbf{z}_{n+1}&=\mathbf{z}_{n}-\frac{\Delta t}{\tau_z}\left( \mathbf{z}_{n}-\bm{\lambda} \right)+\Delta t\, \mathbf{v}_n,\\
	\mathbf{v}_{n+1}&=\mathbf{v}_{n}-\frac{\Delta t}{\tau_v}\mathbf{v}_{n}+\frac{\sqrt{g}}{\tau_{v}}\sqrt{\Delta t}\,\bm{\eta}_n,
\end{align}
where $\bm{\eta}_n$ are standard normal random vectors with zero mean and unit variance. We implemented this algorithm using initial conditions $\mathbf{z}_0=0$ and $\mathbf{v}_0\sim\mathcal{N}\!\left(0,S_v\right)$.


\section{Variogram construction and OUF model fitting}\label{sec:Variogram_plotting}
To construct the variogram and OUF model fit in Fig.\,\ref{Figure:Variance_OUF_Det_Stoch_ICs}B, we use the long-term tracking dataset of lowland tapirs (\textit{Tapirus terrestris}) in the Brazilian Pantanal \citep{Medici_Et_al_2022}. The analysis is performed using the R package, \textit{ctmm} \citep{ctmm_citation}. The ctmm package allows us to analyze animal movement as a continuous-time stochastic process (CTSP) and provides estimates for the relevant movement model parameters. It considers candidate CTSP models such as independent and identically distributed movement (IID), Brownian motion (BM), Ornstein-Uhlenbeck (OU), and Ornstein-Uhlenbeck with foraging (OUF) \citep{Calabrese_et_al_2016}. 

We constructed an empirical variogram from the movement trajectory using the \textit{variogram} function in the  \textit{ctmm} package. The variogram describes the semi-variance in animal location as a function of time lag and provides a diagnostic summary of the temporal autocorrelation structure of the movement path. An initial movement-model guess was obtained by visually matching a theoretical semi-variance function to the empirical variogram using the \textit{variogram.fit} function. This initial estimate was then supplied to \textit{ctmm.select}, which fitted a series of candidate continuous time movement models to the data using perturbative Hybrid Residual Maximum Likelihood (pHREML) and ranked them using the small sample-size corrected Akaike’s Information Criterion (AICc) \citep{Calabrese_et_al_2016}. Model selection ranked the OUF model, which includes both location and velocity autocorrelation, as the best-supported model for the trajectory. We therefore retained the selected OUF model for visualization and parameter estimation. From this model, we extracted the home-range crossing time ($\tau_{z}$, in hours), time scale of directional persistence ($\tau_{v}$, in hours), and spatial variance ($S_{z}$, in $km^2$).


\section{Calculation of the mean encounter rate}\label{app:B}
In this appendix, we summarize the calculation of the OUF mean encounter rate, following the steps in \cite{Martinez-Garcia_et_al_2020}. We start from the definition of the mean encounter rate in Eq.\,\eqref{Equation:Mean_Encounter_rate}, 
\begin{equation}
    \bar{\mathcal{E}}(t) = \int \bar{\mathcal{E}}\big(t\!\mid\!\mathbf{v}_1(0),\mathbf{v}_2(0)\big)\, p_{\rm st}\big(\mathbf{v}_1(0)\big)\, p_{\rm st}\big(\mathbf{v}_2(0)\big)\, d\mathbf{v}_1(0)\, d\mathbf{v}_2(0) = \frac{\gamma}{\pi q^2} \int_{r \le q} f\big(r,t\!\mid\!r_0\big)\, dr,
\end{equation}
where the second equality follows from integrating the initial condition of $\mathbf{v}$ over its stationary distribution, and $f(r,t\mid r_0)$ is the marginal density of the pair separation at time $t$ conditional on the initial separation $r_0$. 

Because the OUF model is a Gaussian stochastic process, the displacement between the two individuals in each coordinate is a Normal random variable with mean equal to the difference between the mean locations of the two individuals. Therefore, $\mu_{\Delta x}(t)=\mu_{x_1}(t)-\mu_{x_2}(t)$ with $y$ following an identical expression. The variance of the displacement is equal to the sum of the variances in the locations of the individuals, $\sigma_{r}^{2}(t)=\sigma_{z_1}^{2}(t) + \sigma_{z_2}^{2}(t)$, where we are already working under the assumption of isotropic movement.

The separation between individuals, on the other hand, follows a non-central chi-distribution. Additionally, if we define a non-dimensional squared distance
\begin{equation}
    u(t)=\frac{\Delta x^2(t)+\Delta y^2(t)}{\sigma_{r}^2(t)},
\end{equation}
with $\sigma_{r}^{2}(t)=\sigma_{z_1}^{2}(t) + \sigma_{z_2}^{2}(t)$, this scaled variable follows a non-central chi-squared distribution with time-dependent non-centrality parameter, 
\begin{equation}
    \hat{\varLambda}(t)=\frac{\varLambda(t)}{\sigma_{r}^{2}(t)}=\frac{\mu_{\Delta x}^2(t)+\mu_{\Delta y}^2(t)}{\sigma_{r}^{2}(t)},
\end{equation}
\noindent The PDF of this non-dimensional squared distance is given by,
\begin{equation}
	f(u(t),\,\hat{\varLambda}(t))=\frac{1}{2}e^{-\left(u(t)+\hat{\varLambda}(t)\right)/2}\,I_{0}\left(\sqrt{\hat{\varLambda}(t)u(t)}\right),\label{Equation:PDF_OUF_Distances_SM}
\end{equation}
where $I_{0}$ is the modified Bessel function of the first kind and order zero. Computing the mean encounter rates thus reduces to evaluating the cumulative distribution function associated with Eq.\,\eqref{Equation:PDF_OUF_Distances_SM} at $u=(q/\sigma_r)^2$, which will return an expression in terms of the Marcum Q-function that is not mathematically amenable for analyzing across parameter regimes \citep{Martinez-Garcia_et_al_2020}. Alternatively, we consider the limit $q\rightarrow0$, which allows us to expand Eq.\,\eqref{Equation:PDF_OUF_Distances_SM} before performing the integral and leads to a mean encounter rate
\begin{equation}
	\bar{\mathcal{E}}\left(t\right)=\frac{\gamma}{\pi q^2}\int_0^{q^2/\sigma_r^2}f\left(u(t),\,\hat{\varLambda}(t)\right)=\frac{\gamma\left[8\sigma_{r}^4(t)+q^2\left(\varLambda(t)-2\sigma_{r}^2(t)\right)\right]}{16\pi\sigma_{r}^6(t)}\,e^{-\frac{\varLambda(t)}{2\sigma_{r}^2(t)}},\label{Equation:Mean_encounter_rate_approx_SM}
\end{equation}
that we can further simplify in the limit where $q=0$ to obtain
\begin{equation}
	\bar{\mathcal{E}}\left(t\right)=\frac{\gamma}{2\pi\sigma_{r}^2(t)}\,e^{-\frac{\varLambda(t)}{2\sigma_{r}^2(t)}}.\label{Equation:Mean_encounter_rate_qequal0_SM}
\end{equation}


\section{Velocity and speed statistics of the OUF model}
\label{app:speed}
The process $\mathbf{v}$ in Eq.\,\eqref{Equation:zdot} is not the animal's velocity, which is given by a linear combination of the animal location and the driving process
\begin{equation}
    \dot{\mathbf{z}}(t) = -\frac{1}{\tau_z}\left[\mathbf{z}(t)-\bm{\lambda}\right] + \mathbf{v}(t).
\label{Equation:velocity_definition}
\end{equation}
\noindent Therefore, the velocity variance combines the variances of the location, $\mathbf{z}$, and of the driving process $\mathbf{v}$ with the covariance between them. In this appendix, we obtain that variance and the resulting speed distribution, which we use as the constraint that stays fixed as $\tau_v$ varies \citep{Bartumeus_et_al_2008}.

We first compute the covariance between the animal location and the OUF driving process $\mathbf{v}$. Writing $\sigma_{zv}(t)$ for the covariance between $z$ and $v$ in one coordinate, and using Eqs.\,\eqref{Equation:zdot}-\eqref{Equation:vdot} we obtain
\begin{equation}
    \frac{d\sigma_{zv}}{dt} = \left\langle \dot{\delta z}\,\delta v\right\rangle + \left\langle \delta z\,\dot{\delta v}\right\rangle
    = -\frac{\sigma_{zv}}{\tau_z} + \sigma_v^2(t) - \frac{\sigma_{zv}}{\tau_v}
    = -\frac{\sigma_{zv}}{\bar{\tau}} + \sigma_v^2(t),
    \label{Equation:sigma_zv_ode}
\end{equation}
where $\bar{\tau}=\tau_z\tau_v/(\tau_z+\tau_v)$ as in Appendix~\ref{app:A}, and the noise makes no contribution because $z(t)$ depends on $\xi(s)$ only for $s<t$. Under the initial conditions of Section~\ref{sec:OUF} the driving process is stationary, $\sigma_v^2(t)=S_v$, and $\sigma_{zv}(0)=0$, so
\begin{equation}
    \sigma_{zv}(t) = S_v\,\bar{\tau}\left(1-e^{-t/\bar{\tau}}\right).
    \label{Equation:sigma_zv}
\end{equation}

\noindent To obtain the velocity variance, we take the variance of Eq.\,\eqref{Equation:velocity_definition} in one coordinate, which gives
\begin{equation}
    \sigma_{\dot z}^2(t) = \frac{\sigma_z^2(t)}{\tau_z^2} + S_v
                         - \frac{2\,\sigma_{zv}(t)}{\tau_z},
    \label{Equation:sigma_zdot_t}
\end{equation}
where the first two terms come from squaring each contribution in Eq.\,\eqref{Equation:velocity_definition} on its own and the third is the cross term, which carries a minus sign because the restoring term appears in Eq.\,\eqref{Equation:velocity_definition} with the sign opposite to that of the driving process.

Because the driving process is stationary, $\sigma_v^2(t)=S_v$, and $\sigma_z^2(t)$ is given by Eq.\,\eqref{Equation:Variance_z(t)}. At $t=0$ both $\sigma_z^2$ and $\sigma_{zv}$ vanish and $\sigma_{\dot z}^2(0)=S_v$, because an animal at its home-range center feels no home-range-center attraction. In these conditions, its velocity is equal to the driving process. As the location relaxes to its stationary distribution, however, the restoring term and the covariance grow, and the velocity variance goes to
\begin{equation}
    S_{\dot z} \equiv \sigma_{\dot z}^2(t\rightarrow\infty)
    = S_v\,\frac{\tau_z}{\tau_z+\tau_v}
    = \frac{S_z}{\tau_z\,\tau_v},
    \label{Equation:S_zdot}
\end{equation}
which follows from Eqs.\,\eqref{equation:moments-pos-stationary} and \eqref{Equation:sigma_zv} evaluated at $t\rightarrow\infty$. Equivalently,
\begin{equation}
    S_z = S_{\dot z}\,\tau_z\,\tau_v,
    \label{Equation:Sz_identity}
\end{equation}
so the home-range size equals the variance of the animal velocity multiplied by the two movement time scales.

In Figs.\,\ref{Figure:OUF_enc_rate} and \ref{Figure:heatmap-threshold} we hold $S_{\dot z}$ fixed as $\tau_v$ varies. Inverting Eq.\,\eqref{Equation:S_zdot} gives the driving-process variance and the noise amplitude that this requires,
\begin{equation}
    S_v(\tau_v) = S_{\dot z}\,\frac{\tau_z+\tau_v}{\tau_z},
    \qquad
    g(\tau_v) = 2\,\tau_v\,S_v(\tau_v)
              = \frac{2\,S_{\dot z}\,\tau_v\left(\tau_z+\tau_v\right)}{\tau_z}.
    \label{Equation:fixed_speed_constraint}
\end{equation}
Under this constraint Eq.\,\eqref{Equation:Sz_identity} makes the home-range size grow in direct proportion to the persistence time scale.


\section{Sign of \texorpdfstring{$\partial\sigma_z^2/\partial\tau_v$ at fixed speed}{dsigma/dtau\_v at fixed speed}}
\label{app:dsigma_dtauv}
Considering the initial conditions used in the main text, the first three terms in Eq.\,\eqref{Equation:sigma_z} vanish as discussed at the end of Appendix~\ref{app:A}, and the variance in location reduces to
\begin{equation}
    \sigma_z^2(t) = \frac{g}{2\tau_v}\,N(t),
    \label{Equation:sigma_from_N}
\end{equation}
where $N(t)$ is the double integral defined in Eq.\,\eqref{Equation:N_definition}. By definition $g/2\tau_v = S_v$, and holding the speed fixed makes $S_v$ depend on the persistence time scale through Eq.\,\eqref{Equation:fixed_speed_constraint}, $S_v(\tau_v) = S_{\dot z}\left(\tau_z+\tau_v\right)/\tau_z$. Therefore,
\begin{equation}
    \sigma_z^2(t) = S_{\dot z}\,\frac{\tau_z+\tau_v}{\tau_z}\,N(t),
    \label{Equation:sigma_fixed_speed}
\end{equation}
and the persistence time scale enters the variance both through this prefactor and through $N(t)$. We consider time $t$ constant and take all derivatives with respect to $\tau_v$ alone. The closed form of $N(t)$ in Eq.\,\eqref{Equation:N_result} is not convenient for this calculation, because it depends on $\tau_v$ through $\bar{\tau}$, $\kappa$ and $A(t)$, and these contributions have opposite signs. We therefore return to Eq.\,\eqref{Equation:N_halfdomain} and change the inner variable to the lag $w = s - s'$,
\begin{equation}
    N(t) = 2\,e^{-2t/\tau_z}\int_0^t\!\! ds \int_0^s\!\! dw\;
    e^{(2s-w)/\tau_z}\,e^{-w/\tau_v}.
\end{equation}
Exchanging the order of integration, so that $w$ runs from $0$ to $t$ and $s$ from $w$ to $t$, and using $\int_w^t ds\, e^{2s/\tau_z} = \tfrac{\tau_z}{2}\left(e^{2t/\tau_z}-e^{2w/\tau_z}\right)$, we obtain
\begin{equation}
    N(t) = \tau_z\int_0^t dw\; e^{-w/\tau_v}\,\varphi(w),
    \qquad
    \text{with}\ \ \varphi(w) = e^{-w/\tau_z} - e^{-(2t-w)/\tau_z}.
    \label{Equation:N_lag}
\end{equation}
This is the same quantity as Eq.\,\eqref{Equation:N_result}, but rewritten so that $\tau_v$ appears in a single exponential. The function $\varphi$ does not depend on $\tau_v$ and is strictly positive on $[0,t)$, because $2t-w > w$ there, and vanishes at $w=t$. Differentiating Eq.\,\eqref{Equation:sigma_fixed_speed} with the help of Eq.\,\eqref{Equation:N_lag}, and noting that inside $N(t)$ only the exponential weight depends on $\tau_v$ so that its derivative produces a factor $w/\tau_v^2$, we obtain
\begin{equation}
    \frac{\partial\sigma_z^2}{\partial\tau_v}\bigg|_{S_{\dot z}}
    = \frac{S_{\dot z}}{\tau_z}\,N(t)
    + \frac{S_{\dot z}\left(\tau_z+\tau_v\right)}{\tau_v^{2}}
      \int_0^t dw\; w\,e^{-w/\tau_v}\,\varphi(w).
    \label{Equation:Derivative_sigma_wrt_tauv}
\end{equation}
Both terms are positive. The first because $N(t)>0$ for every $t>0$, and the second because every factor in its integrand is non-negative and the product is strictly positive on $(0,t)$. Therefore $\partial\sigma_z^2/\partial\tau_v>0$ for every $t>0$ and every $\tau_v>0$. At fixed speed, a higher directional persistence always widens the distribution of animal locations, because individuals hold their heading longer and travel farther from their starting point before turning. Taking $t\rightarrow\infty$ recovers the derivative of the stationary variance in Eq.\,\eqref{Equation:Sz_identity}, $\partial S_z/\partial\tau_v = S_{\dot z}\tau_z > 0$, so the result holds in the stationary regime as well.


\section{Encounter threshold in terms of Bhattacharyya distance}
\label{app:bhattacharyya}

The Bhattacharyya distance between two probability densities is
\begin{equation}
D_B = -\ln \int d\mathbf{z}\,\sqrt{p_1(\mathbf{z})\,p_2(\mathbf{z})},
\end{equation}
where the argument of the logarithm is the Bhattacharyya coefficient \citep{Bhattacharyya1946}. In this appendix we evaluate this integral for the two stationary distributions of animal locations and relate it
to the threshold derived in Section\,\ref{sec:threshold}.

To present the results more generally than in the main text, we assume here the same conditions introduced in Section~\ref{sec:threshold} but allow individuals to differ in their home-range size, $S_{z_i}$. The stationary location of individual $i$ follows an isotropic Normal distribution centered on its home-range center $\bm{\lambda}_i$ and with variance $S_{z_i}$ per coordinate,
\begin{equation}
    p_i(\mathbf{z}) = \frac{1}{2\pi S_{z_i}}
    \exp\!\left[-\frac{\left|\mathbf{z}-\bm{\lambda}_i\right|^2}{2S_{z_i}}\right].
    \label{Equation:stationary_UD}
\end{equation}
The product under the square root is therefore a single Gaussian in $\mathbf{z}$,
\begin{equation}
    \sqrt{p_1(\mathbf{z})\,p_2(\mathbf{z})} = \frac{1}{2\pi S_{z_1}S_{z_2}}
    \exp\!\left[-a\left(\mathbf{z}-\bm{\lambda}_1\right)^2
                -b\left(\mathbf{z}-\bm{\lambda}_2\right)^2\right],
    \qquad
    a = \frac{1}{4S_{z_1}},
    \quad
    b = \frac{1}{4S_{z_2}}.
\end{equation}
Completing the square with
$a|\mathbf{z}-\bm{\lambda}_1|^2 + b|\mathbf{z}-\bm{\lambda}_2|^2
= (a+b)\left|\mathbf{z}-\mathbf{z}^{*}\right|^2
+ \frac{ab}{a+b}\left|\bm{\lambda}_1-\bm{\lambda}_2\right|^2$, where
$\mathbf{z}^{*} = (a\bm{\lambda}_1+b\bm{\lambda}_2)/(a+b)$, the integral over
$\mathbf{z}$ is Gaussian and gives $\pi/(a+b)$. Using
$a+b = (S_{z_1}+S_{z_2})/4S_{z_1}S_{z_2}$ and
$ab/(a+b) = 1/4(S_{z_1}+S_{z_2})$, together with
$\left|\bm{\lambda}_1-\bm{\lambda}_2\right| = R_\lambda$, the Bhattacharyya becomes
\begin{equation}
    \int d\mathbf{z}\,\sqrt{p_1 p_2}
    = 2\,\frac{\sqrt{S_{z_1}S_{z_2}}}{S_{z_1}+S_{z_2}}\,
      \exp\!\left[-\frac{R_\lambda^2}{4\left(S_{z_1}+S_{z_2}\right)}\right],
    \label{Equation:BC}
\end{equation}
and taking minus its logarithm, and writing the sum of the two variances in terms
of their mean $\bar{S}_z = (S_{z_1}+S_{z_2})/2$,
\begin{equation}
    D_B = \frac{R_\lambda^2}{8\,\bar{S}_z}
        + \ln\frac{\bar{S}_z}{\sqrt{S_{z_1}S_{z_2}}}.
    \label{Equation:DB_general}
\end{equation}
The first term compares the separation between the home-range centers with the width of the two distributions, and the second compares the two widths with each other. The second term is non-negative and vanishes only when the two home ranges are equally sized. Specified for the limit case presented in the main text, $S_{z_1}=S_{z_2}\equiv S_z$, the logarithmic term vanishes and the first term in Eq.\,\eqref{Equation:DB_general} is equal to $1/2$ over the threshold  curve $R_\lambda = \sqrt{4 S_z}$.

We can now write the threshold of Section\,\ref{sec:threshold} for individuals with different home-range sizes. From Eq.\,\eqref{Equation:dEdtauv_chain}, the sign of $\partial\bar{\mathcal{E}}/\partial\tau_v$ is set by $\Lambda(t)-2\sigma_r^2(t)$, because $\partial\sigma_r^2/\partial\tau_v>0$ for each individual (Appendix\,\ref{app:dsigma_dtauv}). In the stationary regime $\Lambda=R_\lambda^2$ and $\sigma_r^2=S_{z_1}+S_{z_2}=2\bar{S}_z$, so the sign change always occurs at
\begin{equation}
R_\lambda^{2}=4\,\bar{S}_z,
\label{eq:threshold-general}
\end{equation}
which generalizes Eq.\,\eqref{Equation:threshold} and reduces to it when $S_{z_1}=S_{z_2}$. Inserting Eq.\,\eqref{eq:threshold-general} into Eq.\,\eqref{Equation:DB_general}, the first term equals $1/2$ for any pair of home-range sizes, giving an expression for the threshold in terms of the Bhattacharyya distance of the form
\begin{equation}
D_B^{*}=\frac{1}{2}+\ln\frac{\bar{S}_z}{\sqrt{S_{z_1}S_{z_2}}}.
\label{eq:DB-threshold}
\end{equation}
Because $\bar{S}_z$ and $\sqrt{S_{z_1}S_{z_2}}$ are the arithmetic and geometric means of the two home-range sizes, the logarithmic term measures their asymmetry. It is non-negative by the arithmetic-geometric mean inequality and vanishes only for $S_{z_1}=S_{z_2}$, recovering $D_B^{*}=1/2$. 
\end{appendices}

\end{document}